\documentstyle[preprint,eqsecnum,aps]{revtex}

\begin{document}
\draft
\title{Quenching of $1^+$ excitations in the double giant resonance}
\author{C.A. Bertulani$^{a,b}$, V.Yu. Ponomarev$^{a,c}$ and V.V. Voronov$^c$}
\address{$^a$Institut f\"ur Kernphysik II, GSI,\\
Planckstr. 1, D-64291 Darmstadt, Germany}
\address{$^b$Permanent address: Instituto de F\'\i sica,\\
Universidade Federal do Rio de Janeiro\\
21945-970 Rio de Janeiro, RJ, Brazil, e-mail: bertu@if.ufrj.br}
\address{$^c$Permanent address: Bogoliubov Laboratory of Theoretical\\
Physics, Joint Institute for Nuclear Research, 141980, Dubna, Russia}
\maketitle

\begin{abstract}
The electromagnetic excitation of the two-phonon isovector giant dipole
resonance in relativistic projectiles incident on heavy targets can be
proceed via several intermediate $1^-$ one-phonon giant resonance states. In
two step electric dipole transitions the population of $0^+$, $1^+$, and $2^+
$ two-phonon states are possible. We calculate the amplitude distribution of 
$1^-$ excitations with an RPA formalism, and use it to calculate the
electromagnetic excitation of two-phonon states in second order perturbation
theory and coupled-channels. We show that a conspiracy between angular
momentum coupling and the strength of the electromagnetic fields suppresses
contributions of $1^+$ states to the total cross sections.
\end{abstract}

\preprint{Draft} 

\noindent
Keywords: Two-Phonon Giant Dipole Resonance, Electromagnetic Excitation,
Quenching \\{\it PACS: 24.30.Cz, 25.70.De, 25.75+r} 


\newpage

The electromagnetic excitation of two-phonon states of giant dipole
resonances in heavy-ion collisions at relativistic energies has been studied
in several experiments \cite{1,2,3,4,5}. In LAND measurements \cite{1,2,5}
these cross sections were extracted from total cross sections by separating
a contribution coming from excitation of single dipole and quadrupole
resonances. The observed cross sections are large, on the order of several
hundred of millibarn. However, the first experimental results for the double
giant dipole resonance (DGDR) excitation were found to be enhanced by
factors of 2-3 for $^{136}$Xe \cite{1} and $^{197}$Au \cite{2} as compared
to any theoretical calculations available. Although for the later
measurements in $^{208}$Pb \cite{5} the experiment-theory correspondence is
much better, theoretical calculations still underpredict the DGDR cross
section by about 30~\%. Another part of the problem with double resonances
is related to the experimental position and width of these new resonances
and how they correspond to a harmonic picture of nuclear excitation. For
more details of experimental studies of this problem we refer to reviews 
\cite{rev}.

There are only a few theoretical papers written on this subject. In the
first group the GDR is treated phenomenologically as a single state and the
DGDR is either a sum of these two GDR \cite{6} or a sum of two ``GDR"
phenomenologically mixed with ``GDR$^N$" states \cite{7}. These papers deal
mainly with the problem related to the excitation cross section of the DGDR.
The position and width of the DGDR is a subject for the second group \cite
{8,9,10,11,11a} where the GDR is considered microscopically. In ref. \cite{9}
a fine structure of the GDR was calculated by coupling of RPA 1$^-$ states
to two-phonon configurations and the DGDR was treated as two independent
excitations of the single GDR. In other papers eigen wavefunctions of the
DGDR were used and two-phonon DGDR states were coupled either (a) between
themselves in ref. \cite{8}, or (b) to 1p1h and 2p2h configurations within
second RPA approach in ref. \cite{10}, or (c) to three phonon configurations
in ref. \cite{11}, or (d) to all of them in ref. \cite{11a}. Somewhat aside
from these two groups is the paper \cite{12} where general properties of the
DGDR are treated by means of a sum rule approach.

In the present paper we investigate the contribution of non-natural parity $%
1^+$ two-phonon states to the total electromagnetic excitation cross
sections. The $[1^- \otimes 1^-]_{1^+}$ component of the DGDR was never
considered in previous microscopic studies. They could, in principle, be
responsible for the missing part of theoretical evaluations of the DGDR
excitation cross sections. In phenomenological approaches describing the
single GDR as one collective state, this component of the DGDR is forbidden
by symmetry properties. Taking into account the Landau damping this
collective state splits into a set of different $1^-_i$ states distributed
over an energy interval, where $i$ stands for the order number of each
state. Again, the diagonal components $[1^-_i \otimes 1^-_i]_{1^+}$ are
forbidden by the same symmetry properties but nondiagonal ones $[1^-_i
\otimes 1^-_{i^{\prime}}]_{1^+}$, {\it a priori}, may be excited in two-step
process bringing some ``extra strength" in the DGDR region. Consequently,
the role of these nondiagonal components depends on how strong is the Landau
damping.

To take into account the Landau damping effect we have performed an RPA
calculation for $J^\pi =1^{-}$ states in $^{208}$Pb making use the
quasiparticle-phonon model \cite{13}. The model Hamiltonian includes an
average field for protons and neutrons, treated by a Woods-Saxon potential,
and a residual interaction in a separable form. For open shell nuclei it
also has a term corresponding to monopole pairing. For our RPA calculation
we have used the single particle spectrum and parameters of residual
interaction from ref. \cite{13a}. This calculation provides us with the
spectrum $E_i$ of one-phonon $1^{-}$ states and reduced matrix elements $%
<1_i^{-}||E1||g.s.>$ of their electromagnetic excitation from the ground
state. We produce two-phonon DGDR states with quantum numbers $J^\pi =$~$%
0^{+}$, $1^{+}$ and $2^{+}$ by coupling one-phonon RPA states with the wave
function $|1_i^{-}>_m$, to each other. The index $m$ stands for different
magnetic substates. The wave function of the two-phonon states has the form: 
\begin{equation}
|[1_i^{-}\otimes 1_i^{-}]_{J^\pi =0^{+},2^{+}}>_M=\frac 1{\sqrt{2}}%
\sum_{m,m^{\prime }}(1m1m^{\prime }|JM)|1_i^{-}>_m|1_i^{-}>_{m^{\prime }},
\label{eq:v1}
\end{equation}
for two-phonon states made of two identical phonons while for other DGDR
states it is: 
\begin{equation}
|[1_i^{-}\otimes 1_{i^{\prime }}^{-}]_{J^\pi
=0^{+},1^{+},2^{+}}>_M=\sum_{m,m^{\prime }}(1m1m^{\prime
}|JM)|1_i^{-}>_m|1_{i^{\prime }}^{-}>_{m^{\prime }},  \label{eq:v2}
\end{equation}

In the present calculation we do not include the interaction between DGDR
states, of eqs. (\ref{eq:v1}-\ref{eq:v2}), and we do not couple them to
states with different than two number of phonons as was done in ref. \cite
{11a}. Thus, our two-phonon states $| [1^-_i \otimes
1^-_{i^{\prime}}]_{J^{\pi}}>_{M}$ have excitation energy equal to the sum of
one-phonon energies $E_i + E_{i^{\prime}}$ and are degenerated for different
values of the total spin $J^{\pi}$ and its projection $M$. The reduced
matrix element $< [1^-_{i^{\prime}} \otimes 1^-_{i}]_{J^{\pi}} || E1 ||
1^-_{i} >$ of electromagnetic excitation of two-phonon states, eqs. (\ref
{eq:v1}-\ref{eq:v2}), from the one-phonon state $| 1^-_{i} >_{m}$ is
related, in the boson picture of nuclear excitation, to the excitation of $|
1^-_{i} >_{m}$ from the ground state as follows: 
\begin{equation}
< [1^-_{i^{\prime}} \otimes 1^-_{i}]_{J^{\pi}} || E1 || 1^-_{i} > = \sqrt{%
(1+\delta_{i, i^{\prime}}) \frac{(2J+1)}{3}} < 1^-_{i^{\prime}} || E1 ||
g.s. > .  \label{eq:v3}
\end{equation}
It should be noted that although for the two-phonon states, eq. (\ref{eq:v1}%
), we have an extra factor $\sqrt{2}$, the states of eq. (\ref{eq:v2}) play
a more important role in two-step excitations since they can be reached by
two different possibilities: $g.s. \rightarrow 1^-_{i} \rightarrow [ 1^-_{i}
\otimes 1^-_{i^{\prime}}]$ and $g.s. \rightarrow 1^-_{i^{\prime}}
\rightarrow [ 1^-_{i} \otimes 1^-_{i^{\prime}}]$. Making use of these
nuclear structure ingredients we have performed calculations of DGDR
excitation in relativistic heavy ion collisions in second-order perturbation
theory and with a coupled-channels procedure.

The first effect we observed was that in second-order perturbation theory
the amplitude for this process was identically zero in a semi-classical
approach \cite{14}. This can be understood by looking at figure \ref{fig:1}.
The time-dependent field $V_{E1}$ carries angular momentum with projections $%
m=0,\ \pm 1$. Thus, to reach the $1^+$ DGDR magnetic substates, many routes
are possible. The lines represent transitions caused by the different
projections of $V_{E1}$: (a) dashed lines are for $m=0$, (b) dashed-dotted
lines are for $m=-1$, and (c) solid-lines are for $m=+1$. The relation $%
V_{E1,m=0}\ne V_{E1,m=\pm 1}$ holds, so that not all routes yield the same
excitation amplitude. Since the phases of the wavefunctions of each set of
magnetic substates are equal, the difference between the transition
amplitudes to a final $M$, can also arise from different values of the
Clebsh-Gordan coefficients $(1m 1 m^{\prime}|1M)$. It is easy to see that,
for any route to a final $M$, the second-order amplitude will be
proportional to $(001m|1m) (1m 1m^{\prime}|1M)\ V_{E1,m^{\prime}} \ V_{E1,m}
+ (m \leftrightarrow m^{\prime}$). The two amplitudes carry opposite signs
from the value of the Clebsh-Gordan coefficients. The identically zero
result for the excitation amplitude of the $1^+$ DGDR state is therefore a
consequence of 
\begin{equation}
\sum_{m m^{\prime}} (001m|1m)(1m 1 m^{\prime}|1M)=0~~.   \nonumber
\end{equation}

We have also performed a coupled-channels calculation following the theory
described in ref. \cite{15}. As shown in ref. \cite{15}, the coupling of the
electric quadrupole (isovector and isoscalar) and the electric dipole states
is very weak and can be neglected. We therefore include in our space only
one-phonon $1^-$ and two-phonon $[1^-_i \otimes 1^-_{i^{\prime}}]_{J^{\pi}}$
($J^{\pi} =$ $0^+$, $1^+$ and $2^+$) states. We obtain the occupation
amplitudes by solving the coupled-channels equations 
\begin{equation}
i\hbar a_{\big\{{%
{1^-  \atop [1^- \otimes 1^-]}
}\big\},m} =\sum_{\beta,m^{\prime}}<\Psi_{\big\{{%
{1^-  \atop [1^- \otimes 1^-]}
}\big\}, m}|V_{E1}(t) |\Psi_{\beta,m^{\prime}}> \ \exp \bigg[ i \Big( %
E_\beta - E_{\big\{{%
{1^-  \atop [1^- \otimes 1^-]}
}\big\}} \Big) \bigg] \ a_{\beta, m^{\prime}}   \nonumber
\end{equation}
where $\beta=1^-_i$ or $[1^-_i \otimes 1^-_{i^{\prime}}]$. 
The time dependent electric dipole field is that of a straight-line moving
particle with charge $Ze$, and impact parameter $b$ (we use eqs. (25-26) of
ref. \cite{15}).

Due to the large number of degenerate magnetic substates, to make our
coupled-channels calculation feasible, we have chosen a limited set of GDR
and DGDR states. We have taken six $1^-$ states which have the largest value
of the reduced matrix element $<1^-_i || E1 || g.s.>$. These six states
exhaust 90.6~\% of the classical EWSR, while all $1^-$ states up to 25~MeV
in our RPA calculation exhaust 94.3~\% of it. This value is somewhat smaller
than the 122~\% reported in ref. \cite{18}. It is because the continuum in
our RPA calculation was approximated by narrow quasibound states. From these
six one-phonon $1^-$ states we construct two-phonon $[1^-_i \otimes
1^-_{i^{\prime}}]_{J^{\pi}}$ states, eqs. (\ref{eq:v1}-\ref{eq:v2}), which
also have the largest matrix element of excitation $<[1^-_i \otimes
1^-_{i^{\prime}}]_{J^{\pi}} || E1 || 1^-_i>$ for excitations starting from
one-phonon states \footnote{%
It was demonstrated in ref. \cite{13a,17} that direct excitation of
two-phonon configurations from the ground state is very weak. It allows us
to exclude in our calculation matrix elements of the form $<[1^-_i \otimes
1^-_{i^{\prime}}]_{2^+(1^+)} || E2(M1) || g.s.>$ which correspond to direct
transitions and produce higher order effects in comparison with accounted
ones. These matrix elements give rise to DGDR excitation in first order
perturbation theory. Thus, to prove our approximation we have calculated
such cross sections and got total values equal to 0.11~mb and <0.01~mb for
the twenty one $2^+$ and the fifteen $1^+$ basic two-phonon states,
respectively. These values have to be compared to 244.9~mb for the total
DGDR cross section in the second order perturbation theory.}. The number of
two-phonon states equals to twenty one for $J^{\pi} = 0^+$ and $2^+$, and to
fifteen for $J^{\pi} = 1^+$. The cross section for the DGDR excitation was
obtained by a sum over the final magnetic substates of the square of the
occupation amplitudes and, finally, by an integration over impact parameter.
We have chosen the minimum impact parameter, b= 15.54 fm, corresponding to
the parameterization of ref. \cite{16}, appropriate for lead-lead collisions.

The electromagnetic excitation cross sections for the reaction $^{208}$Pb
(640 MeV/nucleon) $+^{208}$Pb with excitation of all our basic 63 states is
shown in figure \ref{fig:2}. The total cross sections for each multipolarity
are presented in table \ref{tab:1}, together with the results of first-order
(for one-phonon excitations) and second-order (for two-phonon excitations)
perturbation theory. The coupled-channels calculation yields a non-zero
cross section for the $1^+$ DGDR state due to other possible routes
(higher-order), not included in second-order perturbation theory. One
observes a considerable reduction of the DGDR cross sections, as compared to
the predictions of the second-order perturbation theory. The GDR cross
sections are also reduced in magnitude. However, the population of the $1^+$
DGDR states are not appreciable and cannot be the source of the missing
excitation cross section needed to explain the experiments. In general, the
coupled-channels calculation practically does not change the relative
contribution of different one-phonon $1^-_i$ and two-phonon states $[1^-_i
\otimes 1^-_{i^{\prime}}]_{J^{\pi}}$ to the total cross section with given $%
J^{\pi} =$~$1^-$, $0^+$ and $2^+$. But since the $1^+$ component of DGDR,
with its zero value of excitation cross section in the second-order
perturbation theory, has a special place among the two other components, the
main effect of coupled-channels is to redistribute the total cross section
between the $J^{\pi} =$~$0^+$, $2^+$ and $J^{\pi} =$~$1^+$ components.

The calculated cross section in coupled-channels for both GDR and DGDR are
somewhat smaller than reported in experimental findings \cite{5}. This is
not surprising since as mentioned above our chosen six $1^{-}$ states
exhaust only 90.6~\% of EWSR while the photo-neutron data \cite{18} indicate
that this value equals to 122~\%. Due to this underestimate of exhaust of
the EWSR the cross section for DGDR excitation reduces more strongly than
the one for the single GDR. This is because the GDR cross section is roughly
proportional to the total B(E1) value while for DGDR it is proportional to
the square of it. If we apply a primitive scaling to obtain the experimental
value 122~\% of EWSR the ratio $R=\sigma _{(DGDR)}/\sigma _{(GDR)}$, the
last line of our table \ref{tab:1}, changes into 0.096 and 0.101 for the
coupled-channels calculation and for the perturbation theory, respectively.
The experimental findings \cite{5} yield the value $R_{exp}$~=~0.116$\pm $%
0.014. The reported \cite{5} disagreement $R_{exp}/R_{calc}$~=1.33$\pm $0.16
is the result of a comparison with $R_{calc}$ obtained within a folding
model, assuming 122~\% of the EWSR. We get a somewhat larger value of $%
R_{calc}$ (taking into account our scaling procedure) because the B(E1)
strength distribution over our six $1^{-}$ states is not symmetrical with
respect to the centroid energy, $E_{GDR}$: the lower part is enhanced. A
weak energy dependence in the excitation amplitude (see also ref. \cite{7}
for a discussion of this problem), which is also squared for the DGDR,
enhances the DGDR cross section for a non-symmetrical distribution with
respect to the symmetrical one, or when the GDR is treated as a single
state. The effect of the energy dependence is demonstrated for a single GDR
in the top part of figure \ref{fig:2} where the excitation cross sections
are compared to the B(E1) strength distribution. It produces a shift to
lower energies of the centroid of the GDR and the DGDR cross sections with
respect to the centroid of the B(E1) and the B(E1)$\times $B(E1) strength
distribution, respectively. In our calculation this shift equals to 0.26~MeV
for the GDR and to 0.33, 0.28~MeV for the DGDR within coupled-channels and
perturbation theory, respectively.

Of course, this scaling procedure has no deep physical meaning but we have
included this discussion to indicate that the disagreement between
experiment and theory for the DGDR excitation cross sections in $^{208}$Pb
reached the stage when theoretical calculations have to provide a very
precise description of both the GDR and the DGDR to draw up final
conclusions. Work in this direction is in progress now and will be reported
soon.

In summary, we have studied the excitation cross sections of the $1^+$
component of the two-phonon giant dipole resonance in $^{208}$Pb (together
with $0^+$ and $2^+$ components) in relativistic heavy ion collisions. In
second order perturbation theory this cross section equals to zero due to
angular momentum properties. In coupled-channels calculations this component
contributes to the total cross section. But it is appreciably quenched with
respect to other components of DGDR. We indicate that a precise microscopic
calculation of the giant dipole resonance is required to answer the question
whether the problem of enhanced experimental cross section for DGDR
excitation in relativistic heavy ion collisions stands for or not.

\vspace*{10mm}

The authors thank H. Emling for fruitful discussions and suggestions and
acknowledge financial support from the GSI facility at Darmstadt, where an
essential part of this work has been done. C.A.~B. acknowledges the
Brazilian funding agencies CAPES, CNPq, and FUJB/UFRJ for partial support.
V.Yu.~P. and V.V.~V. thank the RFBR (grant 95-02-05701) for partial support.

\begin{table}[h]
\caption{Cross section (in mb) for the excitation of the GDR and the three
components with $J^{\pi} = 0^+, 2^+, 1^+$ of the DGDR in $^{208}$Pb (640
MeV/A) $+^{208}$Pb collisions. Calculations are performed within
coupled-channels (CC) and within the first (PT-1) and second (PT-2) order
perturbation theory, respectively.}
\label{tab:1}\bigskip
\par
\begin{tabular}{|l|c|c|c|}
\hline
& CC & PT-1 & PT-2 \\ \hline
$GDR$ & 2830. & 3275. & 0. \\ \hline
$DGDR_{0^+}$ & 33.0 & 0. & 43.1 \\ 
$DGDR_{2^+}$ & 163.0 & 0.11 & 201.8 \\ 
$DGDR_{1^+}$ & 6.3 & <0.01 & 0. \\ \hline\hline
$DGDR/GDR$ & 0.071 & \multicolumn{2}{c|}{0.075 \hspace*{30mm}} \\ \hline
\end{tabular}
\end{table}

\begin{figure}[h]
\caption{The possible paths to the excitation of a given magnetic substate
of the $1^+$ component of the DGDR are displayed. The transitions caused by
the different projections of the operator $V_{E1}$ are shown by: (a) dashed
lines for $m=0$, (b) dashed-dotted lines for $m=-1$, and (c) solid-lines are 
$m=+1$. }
\label{fig:1}
\end{figure}
\begin{figure}[h]
\caption{The electromagnetic excitation cross sections for the reaction $%
^{208}$Pb (640 MeV/nucleon) $+^{208}$Pb calculated in coupled-channels. It
is shown the excitation of the GDR (top) and the three components $J^{\pi}$%
arbitrary units) over $1^-$ states is shown by dashed lines. For visuality
it is shifted up by 100~keV.}
\label{fig:2}
\end{figure}

\end{document}